# Generative Artificial Intelligence: Evolving Technology, Growing Societal Impact, and Opportunities for Information Systems Research

Veda C. Storey[1] · Wei Thoo Yue[2] · J. Leon Zhao[3] · Roman Lukyanenko[4]

## Abstract
The continuing, explosive developments in generative artificial intelligence (GenAI), built on large language models and related algorithms, has led to much excitement and speculation about the potential impact of this new technology. Claims include artificial intelligence (AI) being poised to revolutionize business and society and dramatically change personal life. However, it is not clear how this technology, with its significantly distinct features from past AI technologies, has transformative potential or how researchers in information systems should react to it. In this paper, we consider the evolving and emerging trends of AI in order to examine its present and predict its future impacts. Many existing papers on GenAI are either too technical for most information systems researchers or lack the depth needed to appreciate the potential impacts of GenAI. We, therefore, attempt to bridge the technical and organizational communities of GenAI from a system-oriented sociotechnical perspective. Specifically, we explore the unique features of GenAI, which are rooted in the continued change from symbolism to connectionism, and the deep systemic and inherent properties of human-AI ecosystems. We retrace the evolution of AI that proceeded the level of adoption, adaption, and use found today, in order to propose future research on various impacts of GenAI in both business and society within the context of information systems research. Our efforts are intended to contribute to the creation of a well-structured research agenda in the information systems community to support innovative strategies and operations enabled by this new wave of AI.

## 1 Introduction

Artificial intelligence is an immensely transformative technology, affecting organizations and individuals in ways that are impossible to fully comprehend without the benefit of a long-term historical hindsight. Artificial intelligence (AI) refers to the ability of computers to perform tasks that have historically required human cognition and other intellectual abilities, such as perception, abstraction, inference, learning, and decision making (Russell & Norvig, 2016). AI is increasingly considered to be pinnacle of human science and engineering (Russell & Norvig, 2016). Industrial estimates are that the global economic value of AI will reach $15 trillion by 2030 (PwC, 2024). The importance of research in AI has long been recognized. It has even been suggested that the country that takes the global lead on AI may emerge as the world's preeminent power (Savage, 2020).

Progress of AI has followed a series of significant paradigm shifts, reflecting the broader trends in technology innovation and diffusion (Perez, 2010). Before the 2000s, the field of AI underwent an "AI winter" due to stagnation in practical advances, during which AI research was primarily dominated by logic-based, model-driven learning, abstraction, and inference methods (Crevier, 1993). Since the 2000s, the ubiquity of large-scale, heterogenous data for training, the continuous expansion of computing power, and the progress in AI algorithm design, shifted the focus almost exclusively toward machine learning (ML) and data-driven AI (Cerf, 2019), where machine learning consists of methods and algorithms that are used to make inferences from data (Goodfellow et al., 2020). With its widespread adoption, ML has begun to rapidly transform organizations and even entire industries.

✉ J. Leon Zhao
leonzhao@cuhk.edu.cn

✉ Roman Lukyanenko
romanl@virginia.edu

[1] Georgia State University, Atlanta, GA, USA
[2] City University of Hong Kong, Kowloon, Hong Kong SAR, China
[3] Chinese University of Hong Kong, Shenzhen, China
[4] University of Virginia, Charlottesville, VA, USA

Researchers in information systems (IS) have always been at the forefront of research on adopting and advancing new technologies to innovate business and social practices. Some examples of past generations of information technologies carefully investigated by IS include organizational data processing, data-driven decision-making and analytics, enterprise resource planning, and electronic commerce. We are now witnessing another AI revolution precipitated by recent developments in deep learning neural networks (Schmidhuber, 2015) and natural language processing (Rotman, 2023). Generative AI (GenAI) refers to AI systems capable of producing content such as text, images, music, programming code and other complex and creative outputs. GenAI is not new. Earlier GenAI included, for example, generative adversarial networks (Goodfellow et al., 2014). However, recent advances in GenAI are driven by the impressive performance of large language models (LLMs). LLMs are computational models that have the capability to understand and generate human language by means of a transformative ability to "predict the likelihood of word sequences or generate new text based on a given input" (Chang et al., 2024, p.2). While LLMs attempt to present human-like abilities, they also face numerous challenges due to rare or unseen words, overfitting, and complex linguistic rules. As such, new LLM architectures and training methods continue to emerge to conquer these and other challenges (Chang et al., 2024).

The result of the advances in GenAI is much interest and speculation concerning the role of artificial intelligence for everyday use, triggered, in part, by the growing popularity of tools such as ChatGPT and Dall-E (open.ai), MidJourney (midjourney.com), Google Bard (bard.google.com), and CoPilot AI (copilotai.com). ChatGPT, is an example of GenAI that can process large amounts of data[1] to create new content (e.g., image, essay, program code, song lyrics). Fundamentally, LLMs make it possible for AI systems to assess and relate to human contexts in relatable and plausible ways (Bender et al., 2021), allowing these systems to preform knowledge-centric tasks they could not accomplish previously. Such capabilities are considered novel, with the potential to support endless new business use cases. Reviews and analyses of GenAI tools, such as ChatGPT, range from "crossing a threshold," where this general tool can be used for a wide range of activities (e.g., new and efficient ways to conduct work) to large-scale plagiarism and misinformation (Savage, 2023; Susarla et al., 2023).

The emergence of GenAI has led to many unanswered questions regarding how society broadly can and should respond to this new technology, particularly concerning its roles and implications. Since the debut of ChatGPT and other GenAI systems, the societal impact of GenAI has led to questions such as: How will GenAI, as technical systems transform business activities? What challenges need to be identified and resolved to ensure productive, ethical, safe, and responsible use of this technology? This gap in understanding the value and risks of GenAI leads to our main research question: *What unique and valuable perspective can information systems research provide with respect to GenAI technology and its impacts on the individuals and organizations?* Since research in information systems has generally been adept at studying the interconnectedness of social and technical aspects of systems (the sociotechnical nature of advances in technology), it should also be relevant for and capable of studying and contributing to the understanding of the value and risks of GenAI.

This paper introduces the concept of a "generative sociotechnical system" to explore what makes GenAI transformative and to support its ongoing diffusion in organizations and society. We develop a theoretical framework that is based upon systems theory (including a sociotechnical lens) and supported by other theoretical foundations (e.g., linguistic theory). The systemist foundation is used to position the components of GenAI and its behavior and helps to develop a research agenda for IS scholarship that builds on the existing strengths and capabilities within our discipline. The use of systems theory also enables us to study how organizations can leverage GenAI technology, how GenAI can be integrated into existing processes responsibly and profitably, and how to effectively couple this new technology with existing organizational resources. In this way, information systems scholarship should focus on exploring the nature and implications of considering GenAI as a component of broader sociotechnical systems, rather than focusing merely on the technical nature of GenAI itself.

This research makes several contributions. First, we propose a theoretical framework for understanding GenAI (Framework for Generative AI as a Sociotechnical System), from a sociotechnical perspective, consistent with existing research in information systems. Second, we show that GenAI is progressing to a point where tools will have the ability to generate unexpected results, requiring boundary conditions and new ways of considering and analyzing research initiatives. Third, we propose the notion of a generative sociotechnical system. By conceptualizing the nature of GenAI from a systems perspective, we can develop a nuanced understanding of GenAI by leveraging existing concepts and theories of IS research. Fourth, we identify fruitful topics for future research.

---

[1] There could be other forms of data, but the basis is textual. Generative AI can "understand" an image because it has seen something similar before. Past images are associated with certain textual data; therefore, new images will also be associated with related textual data.

This paper progresses as follows. Section 2 reviews the advances in artificial intelligence over the past three decades in order to position our understanding of the progression of GenAI. Section 3 presents a theoretical framework for understanding GenAI. Section 4 proposes potential research topics based on the framework and Sect. 5 concludes the paper.

## 2 Related Research

There is a growing body of research within the information systems (IS) and business disciplines that seeks to understand and analyze the benefits and implications of generative artificial intelligence (GenAI) from various perspectives. Dwivedi et al. (2023) adopted a multidisciplinary perspective to detail the multifaceted implications of GenAI for society. Notwithstanding its promise, Fui-Hoon Nah (Fui-Hoon Nah et al., 2023) point to ethical, technological, and regulatory challenges associated with GenAI. Within the IS domain, research has explored the impact of GenAI on teaching (Kajtazi et al., 2023). Jarvenpaa and Klein (2024) discuss the potential of GenAI to assist with theory building in the IS discipline. Susarla et al. (2023) suggest the potential of leveraging GenAI in various capacities to conduct scholarly work. In a broader context, Sabherwal and Grover (2024) argue that the societal impact of artificial intelligence is contingent upon its development and implementation. They emphasize the importance of considering the extent to which AI replaces versus supports humans, integrates physical and digital realities, and respects human limitations. O'Leary (2022) identifies emerging issues associated with large language models (LLMs) generally. Alavi et al. (2024) provide suggestions for how the IS discipline could focus on the role of GenAI from a knowledge management perspective, identifying related research opportunities. We continue these efforts by offering a broad and comprehensive analysis of GenAI as a technology and suggest a fruitful agenda for IS scholarship.

### 2.1 Prior Efforts of Research in AI and Information Systems

To assess the impact of GenAI, as well as the initiatives behind AI and its capabilities, it is crucial to understand the evolution of AI and the technological advancements that have led to its resurgence. This understanding clarifies the possibilities and limitations of AI. Figure 1 illustrates the progression of AI and the significant contributions made by researchers in information systems as AI has progressed from its early work to GenAI. The part above the line shows the major stages of innovations in AI and, below the line, the relevant areas of advancements and related efforts in previous IS research.[2] Figure 1 suggests that information systems research has always attempted to keep pace with research in AI and provides the foundation for this study on GenAI and its applications from the perspective of information systems.

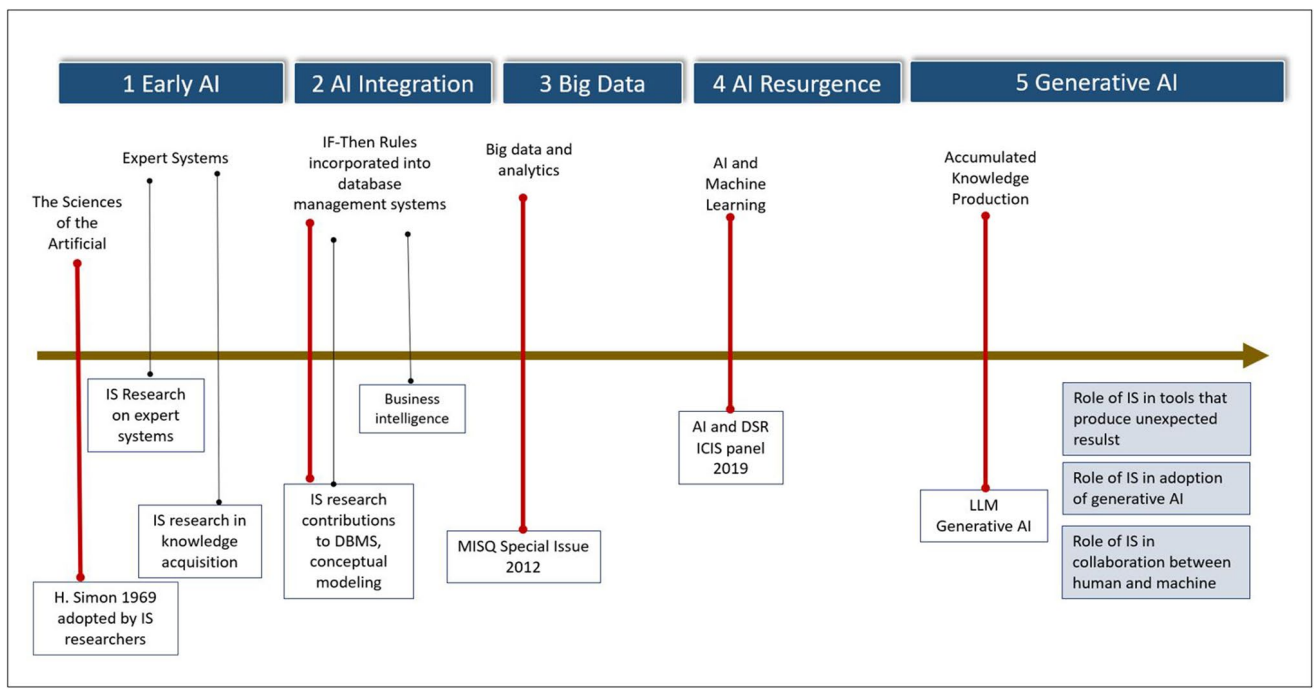

**Fig. 1** Progression of AI and relevant research in information systems

**Table 1** Summary of Early AI Breakthroughs

| Year | AI Artifact | AI Significance | Reference |
| --- | --- | --- | --- |
| 1952 | Checkers | Demonstrated computers can learn to play at level high enough to challenge amateur human player | Samuel (1960) |
| 1955 | Logic Theorist | Proven 38 theorems from Principia Mathematica; introduced critical concepts in artificial intelligence (e.g., heuristics, list processing, reasoning as search) | Newell et al. (1962) |
| 1957 | Perceptron | Birth of connectionism; foundation of Neural Networks (NN), Deep Learning | Rosenblatt (1961) |
| 1961 | MENACE | First program capable of learning to play perfect game of Tic-Tac-Toe | Michie (1963) |
| 1965 | ELIZA | Natural language processing system imitated doctor by responding to questions similar to psychotherapist before conversation became nonsensical | Weizenbaum (1966) |
| 1969 | Shakey the Robot | First general-purpose mobile robot capable of reasoning, integrated research in robotics with computer vision and natural language processing | Bertram (1972) |
| 1969 | The book "Perceptrons" | Highlighted unrecognized limits of feed-forward, two-layered perceptron structure; fundamental shift in AI research to symbolic, disregarding connectionism | Minsky and Papert (1969) |

### 2.2 Evolution from General AI to Generative AI

While GenAI is a recent development, its roots trace back to the very foundations of AI. Initial AI research aimed to create machines capable of human-like understanding and reasoning, with the famous Turing test proposed in 1950 by Alan Turing (Turing, 1950). In 1957, Herbert Simon, John Shaw, and Allen Newell developed a General Problem Solver, intending to simulate human problem solving. In 1959, John McCarthy published a paper with the telling title: "Programs with common sense" (McCarthy, 1959). The development of all-purpose systems that seek to match human cognition across a wide range of tasks later became known as Artificial General Intelligence (AGI) or General AI.

The General AI agenda proved to be less successful than anticipated, despite the high interest and investment levels (Russell and Norvig, 2016). Rather, expert systems, which were rule-based attempts to capture and represent the knowledge of human experts so it could be reused, became the predominant model for AI. The development of General AI was too difficult given the state of computing in the 80s. Rule-based and domain-specific expert systems became the choice of most AI efforts (Segev & Zhao, 1994). With LLMs, the possibility of a General AI became more imminent. For example, the company OpenAI specifically indicates that its product, ChatGPT, can answer questions in any domain (Agrawal, 2023). This suggests that LLMs can be trained on big data that is inclusive of many domains.

To appreciate AI's evolution from its inception to GenAI spanning 70 years of evolution from 1950s (Delipetrev et al., 2020), we recognize three primary development periods: AI Foundations (1950s–1970s); Symbolic AI (1970s–1990s); and machine learning and deep learning (1990s–2020s).

**AI Foundations** (1950s – 1970s). In 1950, Alan Turing published the milestone paper "Computing machinery and intelligence" (Turing, 1950) (Turing, 2012), asking the fundamental question "Can machines think?" Turing proposed an imitation game, known thereafter as the Turing test, where, if a machine could carry on a conversation indistinguishable from a conversation with a human being, it is reasonable to conclude that the machine is intelligent. The Turing test became the first experiment to attempt to measure machine intelligence. The Dartmouth conference in 1956 sparked the start of AI when McCarthy coined the term "artificial intelligence," initiating the emergence of this new scientific field.

The primary mission of the new field of AI was to study "every aspect of any other feature of learning or intelligence [to be] be accurately described so that the machine can simulate it" (Russell & Norvig, 2016). Since then, the AI research community solved problems such as algebraic application problems, language translation, and geometric theorem proving. Table 1 summarizes important AI breakthroughs of that period, based on Delipetrev et al. (2020).

As shown in Table 1, the foundational components of large language models, namely neural networks, deep learning, and national language processing, were initiated in historical projects, Perceptron (1957). ELIZA (1961), and Shakey the Robot (1969), in the foundational period of AI. That is, it took sixty some years from the creation of AI foundations in the lab to the large language models available for general audiences.

---

[2] The authors thank an anonymous reviewer for this positioning.

**Symbolic AI** (1970s – 1990s). The era of symbolic AI focused on the development of systems based on expert-curated rules and application of logic. A major achievement of this period was the development of Expert Systems, which captured expert knowledge and represented it in a symbolic language (Harmon & King, 1985). Expert system research focused on tools for knowledge acquisition to help automate the process of designing, debugging, and maintaining rules defined by the experts. However, domain expert expertise was difficult to obtain, with expert knowledge in constant change, due to variations in regulations and environmental parameters. The period slowly descended into AI winter.

**Machine Learning and Deep Learning** (1990s – 2020). Modern applications of machine learning and deep learning changed the practice of research and development and began accelerating growth across various business, science, and engineering domains. The speed of AI innovation has increased, enabled by big data, expansion of computing power and new algorithms and methods. A major breakthrough was ImageNet in 2009 (Deng et al., 2009), which contained millions of annotated photos in over 20,000 categories and was critical in establishing the legitimacy of using pre-training models to train large deep neural networks. Since then, significant progress has been made in the creation of deep neural networks, which has contributed to advancements in applications such as computer vision, natural language processing (NLP) and robotics.

In 2017, Google researchers developed the transformer architecture (Vaswani et al., 2017), which later became widely adopted in the development of large language models, given its ability to process natural language.[3] The architecture uses self-supervised learning, with a large volume of text corpora from the internet fed into the model to train it. At the heart of the architecture is the construction of self-attention from sentences (Vaswani et al., 2017) that records important information within the context of a sentence on how words relate to each other in different ways. The performance of transformer-based pre-trained models has greatly improved, with large language models distinctively enabled by changing from symbolism to connectionism.

Connectionism gained momentum in the 1990s, gradually displacing symbolism, which was based on a model of the human mind to achieve reasoning similar to that of humans (Zhang et al., 2023). Connectionism is based on connectionist modelling and systems with large networks of extremely simple processors, massively interconnected, and running in parallel, as typically found in various deep learning techniques (Smolensky, 1987). Although Rosenblatt (1961) built a perceptron in 1957—the prototype of an artificial neural network (ANN) – based on the idea of connectionism, neural networks did not achieve broad applications until the 1990s (Zhang et al., 2023).

The current AI breakthrough centers on LLMs and its associated algorithms. LLMs mimic human thinking by means of *connectionism*, which is rooted in machine learning via neural networks, and big data resources gathered from a variety of digital channels. Developing novel AI technologies that are safe, reliable, and extensible, requires a new, explainable and robust AI theory, thereby developing a third generation of artificial intelligence by combining the different AI paradigms (Zhang et al., 2023). Table 2 summarizes significant advances in AI leading to LLMs before 2020. Tables 1 and 2 illustrate selected milestones in the evolution of AI techniques that has led to the most recent boom of Generative AI and LLMs. It is important for IS researchers to be aware that the early AI technology (before 2017) mainly focused on specific applications such as gaming and robots. More recent AI technology (since 2017) has turned to natural language processing and generative applications. However, the concepts and algorithms of the AI technology after 2017 have evolved from the innovative efforts earlier than 2017.

### 2.3 AI Evolution and the Impact on IS Research

The evolution of these AI technologies has significantly impacted both sociotechnical systems and information systems (IS) research. Earlier iterations of AI artifacts were narrow and task-specific, primarily enhancing systems capabilities for specific decision-making processes. For IS researchers, this offered valuable insights into how organizations adapt to technological changes and leverage data to understand various business contexts. However, these earlier systems were limited in their scale and complexity and often lacked the ability to engage with more intricate business operations.

In contrast, the advent of GenAI has broadened its impact across diverse societal and organizational domains. GenAI is more adaptable and context-aware, thus enabling businesses to automate more complex tasks and facilitating deeper insights into decision-making processes. This transformation underscores the critical role of IS research in the adoption and use of GenAI. As highlighted in Fig. 1, IS researchers now face the challenge of exploring how these advanced systems reshape all spectrums of human activities, from organizational work processes to human-AI collaborations, while also identifying unexpected outcomes that may arise from this integration. This paper explores these developments and outlines potential research questions within IS.

---

[3] Transformer-based pre-trained models approach the human-level benchmark with General Language Understanding Evaluation (GLUE), which is based on a collection of English language comprehension problems, rapidly. From the GLUE benchmark, a more rigorous SuperGLUE benchmark test was developed, whereby the models rapidly improved and surpassed human-level standards. From OpenAI's own evaluations, GPT-4 performs exceptionally well on a variety of tests, including reasonings and other examinations (https://cdn.openai.com/papers/gpt-4.pdf).

**Table 2** Summary of Recent AI Advances

| Year | AI Artifact | AI Significance | Reference |
| --- | --- | --- | --- |
| 1989 | Q-learning algorithm | "Learning from Delayed Rewards" improves reinforcement learning | Watkins (1989) |
| 1993 | Solved "very deep learning" task | Scientist solved task with over 1,000 layers in the recurrent neural network (RNN) | Schmidhuber (1993) |
| 1995 | SVM success | Support vector machines applied to text categorization, handwritten character recognition, and image classification | Cortes and Vapnik (1995) |
| 1997 | LSTM architecture | Long short-term memory (LSTM) architecture improved RNN by eliminating the long-term dependency problem | Hochreiter and Schmidhuber (1997) |
| 1998 | Improved gradient-based learning | Combines stochastic gradient descent algorithm with the backpropagation algorithm | LeCun et al. (1998) |
| 2002 | TD-Gammon matched best player | Combines neural nets and Reinforcement Learning (RL) with the self-play method | Tesauro (2002); Blei et al. (2003) |
| 2005 | Stanford robot won award | Drove 131 miles autonomously along an unrehearsed desert track in DARPA Grand Challenge | Thrun et al. (2006) |
| 2011 | IBM Watson won Jeopardy | Watson is Q&A system combining speech recognition, voice synthesis, and information retrieval, among others | Ferrucci (2012) |
| 2012 | AlexNet | Won ImageNet competition, possibly marking inflection point of deep learning | Krizhevsky et al. (2012); Krizhevsky et al. (2017) |
| 2014 | Generative Adversarial Networks (GANs) | Deep neural net architectures composed of two nets; learn to mimic distribution of data to generate content like images, music, speech, etc | Goodfellow et al. (2014); Kingma and Welling (2014) |
| 2017 | Transformer | DL architecture based on self-attention mechanism, important in language modelling, machine translation, and question answering | Vaswani et al. (2017) |
| 2018 | OpenAI Five | Defeated human team at Dota 2, complex video game messier; more realistic than Chess or Go | Pachocki et al. (2018) |
| 2019 | GPT-2 | Large-scale unsupervised language model; can generate coherent paragraphs of text, reading comprehension, machine translation, question answering, and summarization | Radford et al. (2019) |

## 3 Theoretical Framework for Understanding Generative AI

In this section we derive a theoretical framework for GenAI. The framework is intended to help understand GenAI and support proposed research.

### 3.1 Conceptualization of Generative AI

To understand the potential and challenges, it is important to appreciate the mechanisms for processing data inherent in GenAI. To expose these mechanisms, we examine ChatGPT, a major type and exemplar of GenAI. Note that we use ChatGPT as an instance of LLMs but the discussion is relevant to other LLMs. Figure 2 presents a conceptualization of ChatGPT, which we elaborate on next.

As shown in Fig. 2, GenAI, such as ChatGPT is an advanced question-and-answering system based on multiple computational models and big data. ChatGPT and other LLMs take a data-centric approach by training their models to generate human-like responses based on user prompts in a human language. The basic models include transformer (Vaswani et al., 2017), attention ((Luong et al., 2015), GAN (Goodfellow et al., 2014, 2020), reward models (Yu et al., 2020),and RLHF (Li et al., 2023), which are used in generating responses from the texts in the big data repositories (Goodfellow et al., 2014, 2020; Li et al., 2023; Luong et al., 2015; Vaswani et al., 2017; Yu et al., 2020). It is in this sense of text transformation that LLMs are referred to as GenAI. In contrast, in conventional Q&A systems, prestored answers may be matched to keywords in the user questions using some search mechanism, thus being non-generative. Table 3 summarizes the main concepts related to large language models and their mechanisms.

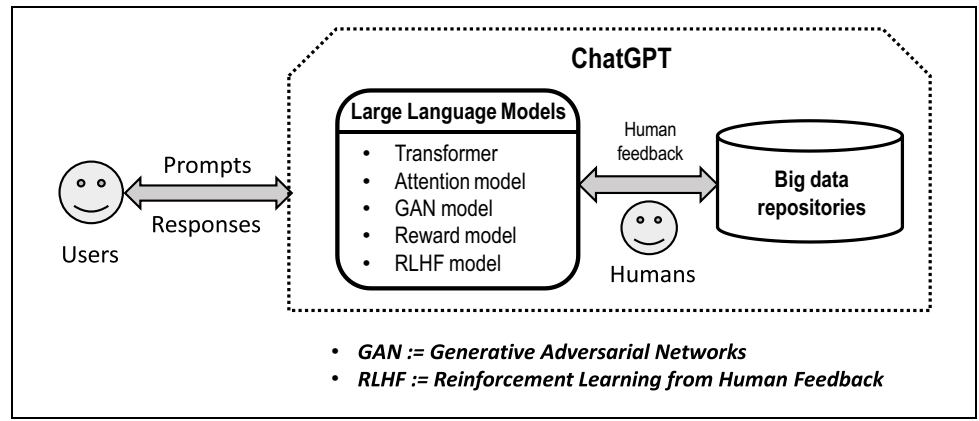

**Fig. 2** Generative AI as advanced Q&A system

We now discuss the main aspects of GenAI advancements as a unique type of information technology. This exploration is not intended to be conclusive but rather to stimulate further discussion and research from the applicational, as opposed to the purely technical, perspective.

**Predictive approach** – ChatGPT is a predictive approach based on deep neural networks. ChatGPT succeeds in natural language processing tasks such as chatting, translation, story making, coding, and other similar tasks in multiple languages. This is because of its representation of human natural language in deep neural networks such as transformers. As in many machine learning techniques, ChatGPT is pre-trained based on existing texts and then used to predict outcomes based on prompt text from the user. This predictive approach means that the responses from ChatGPT are probabilistic and therefore necessarily error prone. This predictive property is easy to comprehend by technical professionals, but requires explanation for non-technical professionals.

**Connectionist models** – ChatGPT is generative in that it creates new connections among words via large language models. As noted, AI evolved into connectionist paradigms after exploring symbolic paradigms and then achieved great success in natural language processing via large language models, leading to ChatGPT and similar tools (Zhang et al., 2023). Neural networks essentially create simple and weighted connections among sentences and words after coding manipulations, allowing machine learning computations on hundreds of billions of inputs (e.g., text tokens). The success of ChatGPT would not have been possible because pure symbolic models are difficult to scale to big data levels. This connectionism can connect ChatGPT naturally with business applications via additional human and machine links to gain efficiency and accuracy. However, the connectionism in ChatGPT does not require deep knowledge of a subject or the intuition or commonsense reasoning of a human. This leaves room for improvements in reasoning and creativity. Thus, although LLMs have shown to produce coherent and human-like text, they do not 'understand' language in the human sense and cannot differentiate between factual from nonfactual information (Saba, 2023).

**Table 3** Fundamental Concepts for Large Language Models

| Concept | Definition | Year |
| --- | --- | --- |
| GAN | Generative adversarial networks (GAN) simultaneously train two models: a generative model that captures the data distribution and a discriminative model that estimates the probability that a sample came from the training data | 2014 |
| Attention | The attention mechanism is used in machine learning and natural language processing to increase model accuracy by focusing on relevant data to enable the model to focus on certain areas of the input data, giving more weight to crucial features and disregarding unimportant ones | 2015 |
| Transformer | Transformers are a type of neural network designed to handle long-range dependencies in text, capturing relationships between words and allowing a model to understand context and meaning across large sequences of inputs. For example, the relevancy and relationship between color, sky, and blue in the question: "What is the color of the sky?" lead to the output: "The sky is blue." | 2017 |
| Reward | Reward modeling is an approach in AI where a model receives a reward or score for its responses to given prompts. This reward signal serves as a reinforcement, guiding the AI model to produce desired outcomes | 2020 |
| RLHF | In machine learning, reinforcement learning from human feedback (RLHF) is a technique to align an intelligent agent to human preferences | 2023 |

**Generic capabilities** – ChatGPT is generic because connectionist models do not distinguish among business sectors or knowledge domains. A single ChatGPT can serve billions of users for questions and answers, without being limited to specific contexts. This feature is a result of its large language models and gigantic neural networks of connections. However, this capability, which is based on a large-scale dataset, also suffers from low accuracy and the possibility of hallucination (that is, making up nonsensical outputs).

**Complementarity with search engines** – ChatGPT results complement those from search engines because the former is predictive; the latter is simply a collection of existing documents relevant to the user query. As such, ChatGPT can be considered a summary of existing documents that are more precisely related to a user query, whereas a search engine simply provides a list of search results to the user. In some way, ChatGPT offers a preliminary result to a user, who may further process the answer by ChatGPT. There is a triangular relationship amongst a user, ChatGPT, and a search engine whereby the user may determine how to use and process the results from ChatGPT and the search engine. This should be a very fruitful research topic on how to manage this knowledge processing ecosystem. The triangular relationship will not replace search engines because the results from ChatGPT might not meet all users' needs, although some users might accept the results, without using the search engine again (Dubin et al., 2023).

**Ability of understanding by ChatGPT** – How ChatGPT comprehends or understands texts in comparison to humans is an important subject of research. Humans learn throughout their lifetime, starting in childhood, and enriching their brains with new knowledge without creasing throughout their entire lifetime. People express ideas, exchange opinions, and create new concepts, words, and theories. ChatGPT can process extremely large amounts of text in existing documents and generate new text, based on user requests. However, it has been noted that AI cannot understand the meaning of a context in the human sense because it does not deal with the deep semantic meaning of the words it processes, demonstrating that there is still a very large gap between artificial intelligence and human intelligence (Zhao, 2022). It is simply not reasonable to assert that ChatGPT can understand or comprehend documents in human sense. There is a significant difference between the generation of new text and creation of new ideas.

### 3.2 Systems Theory: Traditional versus Generative AI Explained

To understand what makes GenAI transformative, we develop a theoretical framework that considers the architecture of GenAI and draws upon systems theory and other theoretical arguments (e.g., linguistic theory). These foundations help to identify the unique characteristics of GenAI and its behavior.

Many efforts to understand and design information technologies have been grounded in systems theory (Chatterjee et al., 2021). Examples include investigations into AI broadly (Skyttner, 2001), trust in AI (Lukyanenko et al., 2022a), and human–robot interactions (Lima & Custodio, 2004). Systems theory has also been previously suggested for understanding ChatGPT (Dwivedi et al., 2023). We hence adopt systems theory to understand GenAI and argue that different types of GenAI (e.g., based on large language models) are types of generative systems.

Systems theory covers related and overlapping theories that deal with the nature of systems, their interactions and uses. For general foundations, we build upon Ackoff (1971) and others, and supplement them with sociotechnical research that studies systems within social contexts (Chatterjee et al., 2021).

A system is a basic scientific and social concept: "an entity which is composed of at least two elements … each of a system's elements is connected to every other element, directly or indirectly. No subset of elements is unrelated to any other subset" (Ackoff, 1971, p. 662). GenAI has numerous inter-connected components. Considering transformer-based LLM, there are encoder and decoder LLMs for the process of input and output systems logic. Conceptualizing GenAI as systems, requires: 1) analyzing key properties of systems, such as emergence; and 2) investigating how this technology can become a component of broader systems, especially sociotechnical systems.

Systems have two kinds of properties (Bunge, 1979, 2018): properties of parts (termed *hereditary)* and properties of the systems themselves (termed *emergent*). For example, the mass of a vehicle is a hereditary property, the sum of masses of its components. Within the context of AI, components of a neural network are the connected nodes. A hereditary property is the coefficient of a connection between nodes.

In addition, systems have *emergent properties,* which are properties of an entire system, rather than of any of its components (Bedau & Humphreys, 2008). These properties emerge when the components become part of the whole and begin interacting with one another in a specific way. Since no component possesses an emergent property, the emergent properties are often not derivable from the knowledge of the properties of the components. For example, solidarity is an emergent property of a political party; no member of the party has this property. This property depends, not only on the beliefs and behaviors of the individual members (as well as extraneous factors), but also on the history, dynamics, and interactions among its members. Similarly, transparency is the emergent property of the entire neural network. While the individual neural network components may be understandable to a human, when these components are put together, they may lack transparency. Similarly, human-sounding outputs of ChatGPT are emergent from the specific connectionist and natural language architecture of GenAI, which, if de-coupled and used separately will fail to render similar types of outputs (Mei et al., 2024).

Emergent properties also shape the emergent behavior of systems, analogous to the *swarming* behavior of a school of fish. As emergent behavior is shaped by the elusive emergent properties, emergent systems are shaped by its emergent properties, and hence its behavior. Emergence leads to qualitative, and, perhaps, ontological novelty (Bedau & Humphreys, 2008). Emergence in systems creates new realities, which do not exist at the level of their components. AI-based systems often can deliver this kind property. For instance, GAN inside GenAI, based on the concept of natural mimicry, can generate images that are deemed novel.

In general, the more *complex* a system, the harder it is to understand and predict its behavior. Systems theory understands complexity as being perceived and actual. Perceived complexity is a human's interpretation and conceptualization of a system as being complex (Li & Wieringa, 2000; Schlindwein & Ison, 2004). The perception of complexity is partially and positively impacted by actual complexity. Actual complexity can be understood as the number of component-parts, along with the way in which these parts are structured and interact with one another (Lukyanenko et al., 2022b). In addition to the general systemic notions, specific considerations apply to particular types of systems (e.g., purposeful, adaptive, organic, artificial, self-organizing, self-reflective, concrete, conceptual) (Ackoff & Emery, 2005). Highlighting the distinctions allows a deeper understanding of GenAI.

Further, we note the distinction between concrete and conceptual systems. Concrete systems are systems made of material (e.g., physical) components (Bunge, 1996). Concrete systems may directly interact with other systems and change as their material components harbor energy, which can respond to and trigger change. Computers and humans are concrete systems. Organizations, which are social systems are also concrete, since they are made of concrete components – humans and their artifacts (Luhmann, 1995).

Conceptual systems are abstract ideas bound together in the mind of a concrete system (e.g., human being) via mental rules, such as logic (Bunge, 1996). Equations, theories, hierarchies, frameworks, language grammar, logically constructed textual narratives (e.g., essays, paragraphs) are conceptual systems. Unlike concrete systems, conceptual systems do not harbor energy and are generated by concrete systems, which expend energy to create, store, modify, and communicate these systems.

Another important type of system is an *adaptive* system. In contrast to hard-wired or rigid systems, adaptive systems are capable of responding to environmental changes by reconfiguring their internal states. A subset of such systems are *complex adaptive systems,* defined as "systems composed of interacting agents described in terms of rules. The agents adapt by changing their rules as experience accumulates" (Holland, 1992, p. 10). Complex adaptive systems include natural organisms and artificial systems, such as complex machinery, including those based on AI. Such systems commonly rely on feedback loops, wherein the outputs of the system become its inputs, and hence can modulate or amplify the system's behavior. An important consequence of adaptivity is increased difficulty in anticipating and predicting the behavior of such systems (Holland, 1992).

The objective of a GenAI process is to produce novel, complex, and self-contained outputs, in contrast to traditional machine learning, which focuses primarily on learning decision boundaries based on patterns extracted from data (Walters & Murcko, 2020). There is a profound qualitative difference between GenAI and traditional, discriminative type systems. Traditional AI technologies are fundamentally decision models, mathematical structures that seek to connect inputs to outputs in a relatively straightforward manner. Although opaque and complex, their operating principles resemble that of a measuring tape or a calculator. Often, the solution or output space is bounded, making them focus on specific tasks, therefore considered to be narrow AI.

In contrast, GenAI, which is also based on billions of iterations of training, and expressed as complex mathematical structures, determines the output in a much less straightforward way. Rather than rules that connect input to output, GenAI is a set of parameters that guides the development of a self-contained output based on an input request. A useful analogy is natural language systems.

A prevailing view in linguistics is natural languages are guided by universal grammar (UG) – the instinctive principles are instantiated when a speaker learns a particular language and uses it (Chomsky, 1986). The UG sets general principles and parameters (e.g., every word can be identified with a linguistic category). Hence, UG is a meta-language. The result of UG principles and parameters is the generativity of human languages. For a goal, infinitely different expressions can be generated. Also, following UG, the outputs of natural language are coherent and self-contained. For example, a sentence, a paragraph, or an essay all have internally consistent and coherent structure, guided by the principles and parameters of UG.

In much the same way as natural language and other generative systems (e.g., number system), GenAI can produce potentially unlimited outputs, based on similar inputs (under the common assumption that the parameter of *temperature* is above zero). While these outputs are ultimately grounded in the data, novelty is in the new connections and transformation of these data. Given the extremely large set of parameters and data sources, it is nearly impossible to predict every possible way that GenAI may connect an input to an output. Although the comparison of tools, such as ChatGPT, with human creativity remain controversial, GenAI promises significantly higher level of output novelty compared with most other forms of AI.

While this may suggest it is impossible to fully control GenAI, systems theory and linguistic theory offer insights on how to potentially predict and manage these types of technologies. Specifically, the principles and parameters of GenAI are important. The linguistic structures of a language are learned by the system, mapped to another language or images, and eventually applied for different tasks; for example, a text instruction to generate a Python script.

Due to their generative ability, AI systems, such as ChatGPT, exhibit much greater complexity than traditional machine learning systems (Bender et al., 2021). These systems are also capable of relating and combining data types that go beyond text (e.g., image and speech), and, as a result, are expected to have a significantly larger number of emergent properties, leading to a variety of emergent behaviors. Following systems theory, the ability of AI systems to predict behavior (relationship of inputs to outputs) in generative systems), is considerably reduced as compared to traditional machine learning systems.

Systems theory permits other notable insight into generative systems. Not only is GenAI a system itself, but its inputs and outputs are often systems (not just system components). For example, text-to-image systems, such as DALL-E, combine LLMs and diffusion models. The prior constructs the understanding of images connected to the input texts; the latter follows a probabilistic model to synthesize images similar to the input texts. The two models were previously applied independently in systems focusing on respective tasks.

Previous generations of AI operated with only system components as their inputs and outputs. A typical machine learning model accepted inputs of a particular predefined format (controlled extraneously by the user interface), and generated outputs, which, strictly speaking, were not usable by themselves, but required interpretation and integration into larger (conceptual or concrete) systems. For example, a machine learning model could generate a credit risk for a customer, based on a predefined feature vector corresponding to parameters of a particular customer. However, the results such as $<353, 0.8>$, must be interpreted and integrated into a larger conceptual system.

Generative AI, however, can accept and generate *component systems*, as well as *systems in their own right*. Indeed, when ChatGPT writes an essay, creates a Python script, or answers a question; when DALL-E or MidJourney paints an image; or when MusicLM creates music, the result is a standalone *conceptual system*. This system no longer needs to be embedded in another system to be usable. It could be directly used, for example, to listen to music, enjoy an image, or submit an essay as a class assignment (which, of course, poses fundamental ethical and pedagogical questions).

Finally, the ability to produce such a wide range of outputs, both component systems as well as systems in their own right, allows for a much greater diversity of the utilization of GenAI within organizations. As previous artificial intelligence, GenAI can be embedded in other technologies by harnessing its components system outputs in order to derive a coherent broader system. For example, an existing decision support system within an organization can consult GenAI for some tasks and present an output to a user which is in part based on the responses from a GenAI technology. In addition, GenAI can be integrated more directly within organizational structures, because it produces standalone conceptual systems. For example, organizational employees can use this technology in order to prepare reports, write essays, or conduct a review of a domain. Given these diverse capabilities, an important question is how to appropriately leverage this *systemic versatility* of GenAI.

Based on systems theory and linguistics outlined above, we propose that GenAI has three generative properties that distinguish it from the machine learning and artificial intelligence technologies of the past.

- **Strong Emergence**. Whereas previous AI systems might possess emergent properties, GenAI has a particular significant ability to behave in a manner that is not directly derivable from the properties of its components and is very distant from them. This is because the outcomes of GenAI are the result of transformations from the prompts in combination with the complex knowledge of a GenAI system. Strong emergence can have both positive (e.g., ability to generate creative content) and negative consequences (e.g., difficulty to control GenAI and assure it does not harm or disadvantage people).
- **Generative Novelty**. Rooted in strong emergence is generative novelty. Generative AI has the capacity to produce both expected and unexpected outputs, based on a given input. The outputs are products of billions of parameters tuned through billions of iterations over big training data. While these outputs are ultimately grounded in training data, the novelty lies in new ways to transform data and identify unseen patterns.
- **Systemic inputs and outputs**. Generative AI has the ability to accept and produce coherent, self-contained outputs (such as self-contained responses, essays, images, animations, music). Effectively, GenAI produces self-contained conceptual systems (as opposed to snippets of decision rules), along with system components. That is to say, the systems as outputs are mapped from the system inputs based on the sophisticated transformation algorithms.

These three properties lead to new opportunities for human–computer interactions. Their organizational effects reshape the possibilities for IS research in ways that are unprecedented in the history of information systems.

### 3.3 Sociotechnical Perspective

Research in information systems is highly focused on *sociotechnical* issues where technical components are considered within the context of broader social systems (Sarker et al., 2019). Because the sociotechnical perspective fundamentally originates in the systems theory, we used to understand the nature of GenAI, it is reasonable to consider GenAI from the *sociotechnical* perspective.

Figure 3 presents a *Framework for GenAI as a Sociotechnical System*. Because it is based on the systems perspective and follows the sociotechnical perspective, we suggest that IS research should consider the technical properties of GenAI technology within the context of other systems, organizations, individuals, and processes. Specifically, the distinct foundational properties of GenAI of strong emergence, novelty and systemic inputs and outputs, produce new challenges and opportunities for investigating the design, use and impact of this technology in organizational and societal contexts. Such investigation permits information systems users to adopt and leverage the extensive existing knowledge of organizations and systems and to integrate this knowledge with the new fundamentals of GenAI. For example, with extensive information systems research on e-commerce and personalization, the integration of GenAI in e-commerce platforms (e.g., the Bing search engine) creates opportunities to investigate: the role of generative artificial intelligence in user adoption; trust towards e-commerce technologies; and the ability to facilitate commerce and information exchange. Because of the three fundamentally new properties of GenAI (See Fig. 3), this technology can be coupled in novel ways with the broader systems. This can lead to a research opportunity to study the intersection of the technical aspects of GenAI and the ways in which these capabilities succeed or fail to support, augment, and enhance the systems they become part of, or interact with.

A sociotechnical system based on GenAI should behave consistently with human values (AI alignment) (Hagendorff & Fabi, 2022). The alignment involves precisely prioritizing fundamental human values within the LLM. To represent these values in an AI system, humans are often involved in fine-tuning an LLM using reinforcement learning from human feedback (RLHF). Researchers in information systems should consider the broader presence of a system, including the boundary conditions that define its presence and impact within the context of an individual, organization, or society.

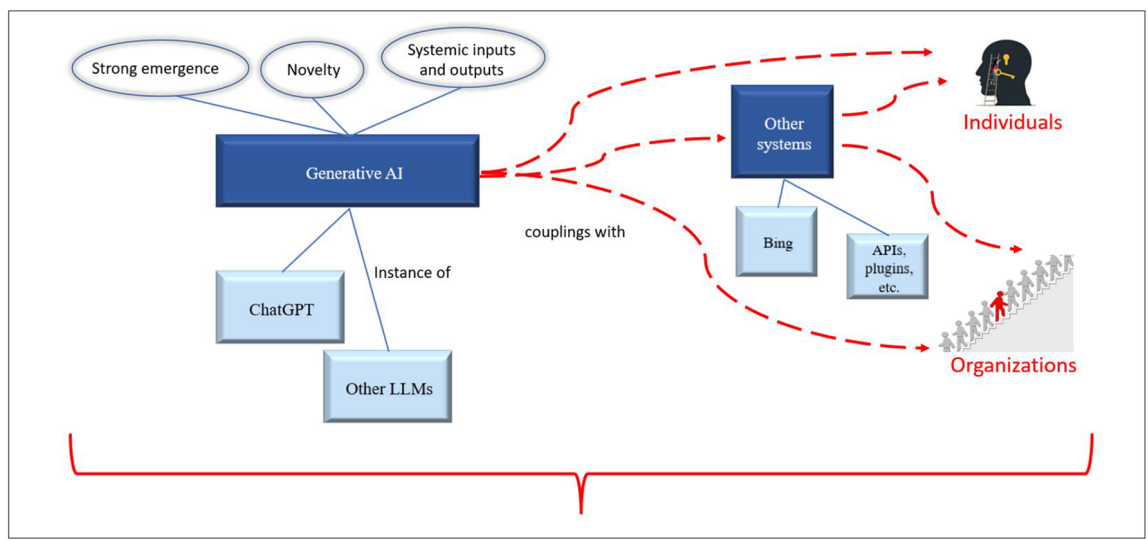

**Fig. 3** Framework for Generative AI as a Sociotechnical System

Adopting the sociotechnical perspective, we suggest that there are broad implications of how GenAI has the potential to transform organizations and society. The first is the *disruption and impact* on applications that will dramatically improve human and organizational productivity. As GenAI technology advances lead to new possibilities, it will assist in the development of creative and innovative solutions. Second, a GenAI system will foster closer *human and systems relationships* that span both personal and business spaces. This leads to research issues focused on human-AI collaborations. Third, the versatile nature of GenAI presents numerous uncertainties, some of which may provide new opportunities but also pose threats to individuals, organizations, or society, which we consider to be the dark side of GenAI. Generative AI systems, thus, require a *new design paradigm*. Potential topics of designing systems with GenAI include societal concerns related to job automation and displacement of human workers, which can lead to demands for new skill sets and retraining.

## 4 Research Opportunities for Information Systems Research

This section proposes research themes and topics for future research on GenAI for information systems and related disciplines. They build on the sociotechnical theoretical foundations that prioritize the intersection of the technical properties of GenAI and its social entities (individuals, organizations, society).

### 4.1 Disruption and Impact

Potentially, GenAI could substantially enhance individual and organizational productivity, causing disruptions in numerous industries. IS researchers have a long track record of studying such impacts of new technologies.

**Productivity Enhancement and Task Automation** Machine learning algorithms automate data classification and prediction tasks, allowing more efficient data processing and analysis. Robotic process automation can automate routine business processes, such as data extraction and data integration. Because GenAI can process language, it can automate a wider range of jobs, leading to new use cases. For example, ChatGPT is affecting advertising because of how well it can prepare marketing and advertising messages.[4] AI systems can free up people to perform more fulfilling or creative work. This can lead to better efficiency, faster decision-making, and higher total productivity. IS researchers have traditionally aimed to quantify the value of technology under various circumstances (Chang & Gurbaxani, 2012). Similarly, there is a need to assess the value introduced by GenAI within the greater business context.

**Transformation of the Workforce and Organization** The adoption of GenAI may necessitate a transformation of the workforce (Alavi, 2024; Alavi et al., 2024). Some job roles and even occupations may become obsolete or less relevant, with new roles and skills in demand. To recognize the value of AI, for instance, organizational change may be required (Reis et al., 2020). The role that senior management plays in the organization is also of interest to researchers. The capacity of GenAI to automate and optimize business processes has the potential to revolutionize businesses. Using machine learning and natural language processing can streamline operations and decision-making (Kanbach et al., 2023). Therefore, IS researchers should continue to address questions regarding organizational technology adoption.

**Governance and Policy** Implications. Given the disruptive potential of GenAI, the rules of engagement with new technology may change. There are many issues specific to IS, such as defining the governance model of new technology that may possess very different characteristics than those of traditional systems. This can affect an organization's strategic orientations for managing AI-driven systems. Included are research opportunities involving use case scenarios, stakeholder engagements, and risk management approaches. Governance and policy mechanisms will invariably influence organizational performance (Wu et al., 2015). Moreover, IS scholars have contributed to understanding the interactions of these mechanisms in system's use with the presence of internal and external organizational factors (Xue et al., 2021). Significant research is required to study the impact of systems built with the support of GenAI on organizational agility and strategy. Empowered users and IT departments supported by GenAI should be able to develop and maintain information technologies at a faster pace than in the past. This raises the question of what organizational capabilities are needed to match the expected increase in information systems adaptation and evolution. IS scholars can collaborate with business communities to study IT-strategy alignment, organizational agility, and flexible organizational routines and processes (Feldman & Pentland, 2003).

### 4.2 Human Machine Collaboration

Human-AI Collaboration is leading to new research issues related to fairness, unintended consequences, and regulations for human–machine collaboration. Although many

---

[4] https://www.forbes.com/sites/bernardmarr/2023/01/17/how-will-chatgpt-affect-your-job-if-you-work-in-advertising-and-marketing/?sh=7c62f73f39a3

academic disciplines study human-AI collaboration, information systems have a natural capability and responsibility to help make progress in areas such as: design of human-AI systems; human and AI behaviors; economic issues of AI-assisted businesses and societies; and organizational issues of AI-intensive firms. Sample topics in human-AI collaboration can include the following:

**Human and AI Agency** Generative AI systems may facilitate humans to be the delegator; systems play the role of supporters or enablers. Although this role reversal has been hypothesized previously (Demetis & Lee, 2018), GenAI dramatically increases its scope and scale. We must define and explain the AI and human responsibilities in collaboration, including task responsibility, delegation, and performance evaluations. As systems mature, they may also be able to carry out the owner's plans (or agendas). Unlike past recommendation systems, the emphasis of new AI systems might be on trying to change what people like and making a corresponding recommendation. Contexts, user engagement, recommendation systems, and systems adoption are often considered separately, but a more comprehensive consideration from IS researchers will be required.

**Human-AI Interface** For a user, emerging challenges, such as prompt engineering, are interface-related issues (Zamfirescu-Pereira et al., 2023). Traditionally, interface design challenges have been concerned with making human use easier. This might be accomplished by structuring the interfaces based on predetermined choices or a flexible format (Lukyanenko et al., 2019). Making a selection from an existing list is more straightforward than typing text. Future research can investigate design approaches for GenAI systems that emphasize more open, flexible and dynamic communication process between humans and systems. However, these communication processes are dependent on how the LLM is formed, which is opaque to users.

Despite being trained with text that could reflect human values, prompt communication does not share the same communication process as humans. For example, AI systems often require more specific contextual information to accurate responses because they do not have the same reasoning ability as humans. Studies on prompt engineering could be aimed at determining the best input approaches that could be fed into the systems to interface with humans more effectively (Yao et al., 2023). Furthermore, not all humans will be comfortable with, or know how to pose the right questions to obtain the desired answers. The generative aspect of AI, could shift the conversation into topics not desired or wanted by users, followed by misconstrued prompts. Research is required on how to manage the openness and fluidity of the system-human interaction to ensure positive outcomes for users and organizations.

**Human-AI Configuration** Information systems researchers have rich experiences in the analysis and design of business processes in the past, leading to various models of process and workflow design (Stohr & Zhao, 2001). In the era of GenAI, process modeling will need to consider the increasing use of AI agents in place of humans, referred to as human-AI configuration (Berente et al., 2021). This will require the development of theories and principles of system design based on the understanding of human and AI behaviors in a collaborative environment where human and AI agents share information and responsibilities under emerging regulations, governing acting human agents and other humans responsible for developing and controlling the AI agents (Park et al., 2019). There will be a new area of research and development with interesting and challenging issues. It is likely that this human-AI configuration will require interdisciplinary expertise from multiple fields, such as artificial intelligence, information systems, computer science, software engineering, industrial engineering, management and economics, and law.

**Design principles of human-AI systems** The study of human-AI collaboration will likely lead to theories and guidelines that will be instrumental in the analysis and design of AI-intensive systems in business (Fügener et al., 2021). There will be multiple aspects of design beyond human-AI configuration. As such, principles of design, development, and application of human-AI systems will be needed to facilitate the training of new talent in related disciples, development, and implementation of AI-intensive systems in business, and the creation of a new branch in the IT sector. These design principles will require new theories in human behaviors in the presence of AI agents, since the benefits and constraints of adopting AI agents will have an impact on human decisions in human-AI systems (Wang et al., 2019). The likely result will be research topics such as: mixed behavior under human and AI collaboration; reward and risk of decision making in human-AI systems; and process management in human-AI collaboration. Thus, the emergent applications in business are resulting in more attention on the research area of human-AI collaboration (Sowa et al., 2021).

### 4.3 Dark-Side of Generative AI

Throughout the development of information systems, researchers and practitioners have expressed concern about the unintended consequences of GenAI systems (Tarafdar et al., 2014). It is commonly believed that technology is neutral and its impact is determined by how users employ it (Yue et al., 2019). However, Generative AI is a highly adaptable technology. For people with devious purposes, there could be numerous negative consequences.

**Intellectual Property Right Infringement** The potential for infringement of intellectual property rights is a significant challenge. The technology has the capacity to generate novel content based on copy-protected data. For example, it has rapidly created music that imitates the style of artists, as well as voice covers that imitate their voices precisely. This concerns human artists, whose artistic talent could be "cloned" or stolen, thereby interfering with their ability to monetize their work (Yeshchenko et al., 2019). Such technology can lead to ethical and trust issues and can facilitate the creation of "digital twins" of deceased individuals, complete with their voices and communication styles (Basilan, 2023), again raising ethical concerns.

**Misinformation and Deepfakes** The creation of false digital information, such as images, content, and voice, that appears authentic, but is actually fake, is concerning. These deepfakes have the potential to be used for misinformation or fraudulent purposes by malevolent actors (French et al., 2024). This necessitates oversight, corrective measures, and governance (French et al., 2024). Similar to the impact of email technology on the proliferation of SPAM, the democratization of the technology has lowered the barrier to creating deepfakes, resulting in an abundance of content, making it difficult to differentiate authentic and fake information (Haidt & Schmidt, 2023).

**Emotional Manipulation and Deception** Programs can produce responses that appear human-like and might even be mistaken as showing personality, which can influence more advanced forms of communication, including intentions. For example, technology could be designed to keep people on websites and encourage them to buy something, by recommending products or services based on what people might like or need. This scenario has been assumed in prior IS scholarship (Qiu & Benbasat, 2010). The intuitiveness and human-like responses of GenAI, however, could give humans a deceptive sense of emotion, presence, and consciousness, possibly leading to deeper emotional connections and allowing for manipulation and deception. There could be issues associated with whether minors are emotionally ready to engage these tools (Kelly, 2023).

**Hallucination and Biases** Despite many impressive capabilities, GenAI suffers from several additional shortcomings and limitations. Biases and hallucinations, in particular, have been identified as an especially difficult issue to overcome. The "hallucination" problem for large language models (Ji et al., 2023) means that GenAI has a tendency to generate nonsensical outputs based on specified inputs. Related challenges are biases, which can stem from using data that does not represent reality or systematically distorts it, or improper training with the data, which may introduce additional or compound existing distortions. Tools such as ChatGPT, have been shown to exhibit a variety of such biases, including political, moral and cultural (Motoki et al., 2024). Hence, modern GenAI appear to be the biased towards the Democrats in the US, Lula de Silva in Brazil, and the Labour Party in the UK (Motoki et al., 2024). If left unaddressed, such biases can result in erroneous decisions based on these tools and may undermine trust in particular GenAI tools, their vendors, or the entire artificial intelligence industry.

Regardless of its impressive performance, GenAI lacks the human ability to understand the meaning of its inputs (e.g., prompts of ChatGPT) and outputs (e.g., essay written in response to the prompt). Generative AI is driven by statistical probabilities of words and, more generally, pattern co-occurrences, irrespective of their actual real-world meaning (Bender et al., 2021). Large language models are merely based on probabilities of a particular word, or sentence (or more generally, token) being appropriate for particular context. Unlike humans that have imported experiences, the tools fundamentally lack ability to relate these tokens to human feelings, and thoughts and bodily experiences. The result might be potentially dangerous recommendations, because the technology lacks the ability to understand the human context around these recommendations (Storey et al., 2022).The sociotechnical lens of our paper suggests a perspective on investigating the hallucination and biases issue. On the one hand, design science researchers can contribute to the interventions that ameliorate these biases by conducting technical research. For example, data management scholars can investigate ways to evaluate biases in existing data and devise ways to procure additional data to make it more representative. Machine learning scholars can contribute by developing algorithms for handling bias to data through statistical techniques and additional processing. On the other hand, researchers should study the impact of hallucination and biases on the ways GenAI is used by people and is integrated into organizational routines. For example, an important research opportunity is understanding the boundary conditions for using GenAI for mission critical tasks. Similarly, an important research question is how to mitigate some of these biases by using additional, supporting technologies, which may be free of hallucinations and biases (such as vetted knowledge databases).

**Energy Usage and Environmental Impact** Another issue is that GenAI tools demand a great deal of resources (energy). Machine learning is having a staggering environmental footprint (Wu et al., 2022): the cost to train ChatGPT-3 was approximately 936 MWh, enough to power close to 100 homes for a year.[5] Information systems scholars can investigate the technical solutions to the energy impact problem, by devising algorithms that minimize energy usage. At the same time, a major research opportunity is understanding the full, immediate and delayed impact of widespread usage of these technologies on the environment.

---
[5] https://medium.com/mlearning-ai/an-ai-model-that-is-energy-efficient-is-just-as-important-as-its-purpose-71d17822a183

**Transparency** As discussed, GenAI appears to be significantly less transparent than the traditional AI approaches of symbolic models. Large language models contain billions of parameters making it impossible for humans to fully understand how they make their decisions (Bender et al., 2021). Transparency of AI has long been a major barrier for trust and organizational adoption of AI technology (Bedué & Fritzsche, 2022). It is incredibly challenging to understand the logic behind LLMs, which raises significant concerns about potentially undetected biases, trust, and reliance on these systems, especially in sensitive applications (Kaneko & Baldwin, 2024). An important research opportunity is investigating the extent to which relying on such fundamentally opaque models may be unsafe, especially in particular scenarios.

The rapid adoption and incorporation of AI technology in society can result in the transformation or extinction of industries and occupations, causing workforce disruptions. It will be necessary to understand the potential repercussions to ensure that the largely opaque AI technology is aligned with human values and interests.

### 4.4 Designing Systems with Generative AI

Information systems were created to increase organizational efficiency and effectiveness by augmenting productivity and automating processes across various organizational functions (Zuboff, 1988). Generative AI will undoubtedly accelerate this trend. Moreover, this new technology solidifies the cloud paradigm, which promotes collaboration and integration of IT operations and services (Rajput, 2023; Seseri, 2023). Generative AI is being applied to information systems development for tasks including writing programming code (e.g., Python, JavaScript, HTML or VBA) and producing digital artifacts or components of information systems (e.g., videos, images) (Gewitz, 2023).

**Information Systems Development** A challenge is the rethinking of the role of users as developers. The ability of GenAI to develop systems raises fundamental questions about the process of information systems development. The redefined nature and role of end-users as developers is likely to accelerate the already existing trend where non-IT professionals significantly increase their engagement with IT development (Legner et al., 2017). A new kind of user, "empowered users," refers to the diverse and heterogeneous group of non-IT professionals who are motivated to take autonomous initiative and action to implement a desired change using information technology. Organizational non-IT employees increasingly implement their own solutions, such as workarounds to existing systems (Alter, 2015) and completely new solutions, especially in areas such as analytics (Khatri & Samuel, 2019). User empowerment is occurring, not only within business organizations, but also within society broadly, as people leverage GenAI's capabilities to develop websites, apps and digital media. This bottom-up development by empowered users challenges the relationship between IT and users (Chua & Storey, 2016), leading to new avenues for research on how to best support such users.

**Implementing Generative Systems** The development model shifts the traditional human-centric business process to a human-in-the-loop process. There are, thus, research opportunities to redefine the role of humans and systems in the business process. Generative AI gives new impetus to the notion that a human is becoming an artifact, shaped by IT (Demetis & Lee, 2018). The systems are multi-model in nature, which is closer to how humans operate.

Many research opportunities exist in knowledge representation in the synchronization of text, image, video, and sound, based on business tasks. The transformation of different data forms to knowledge is made possible by GenAI systems, but requires new ideas of how businesses bridge proprietary data with the new technology. Another question is the boundary conditions of systems development with the support of GenAI. Some industries, such as manufacturing, construction and entertainment, have already become users of automated systems development (Seidel et al., 2018; Verganti et al., 2020). An important opportunity is understanding for which settings and industries GenAI-supported development can be effective, and where it may result in net negative outcomes. For example, design automation with GenAI may pose increased risks in mission critical and highly sensitive environments, as well as heterogeneous environments. An important design issue is how to take advantage of the capabilities of GenAI in the settings, while ensuring the safety, comfort and well-being of those affected by these designs. Greater human involvement and human-in-the loop in these contexts could be one of the solutions.

**Data and Technology Integration** The rapid explosion of data from multiple, heterogeneous sources presents a significant challenge. Traditionally, data integration focused on matching data from well-structured sources (e.g., database schema mapping (Batini et al., 1986)). However, the approach has limitations in ensuring the completeness and accuracy of responses. Combining disparate sources from different domains with varying levels of quality requires integrating additional techniques and technologies to fine-tune the AI systems' performance. Retrieval augmented generation (RAG) offers a promising approach by providing LLM with additional domain-specific knowledge to improve output performance (Ke et al., 2024). Similarly, in-context learning (ICL) provides LLMs with relevant examples and instructions through prompts. This approach instills a chain of thought during response generation, leading to a system's ability to coherently utilize information and knowledge learned in real time (Tang et al., 2023). Researchers in information systems should focus on research questions that explore how to most effectively implement these augmenting methods in knowledge production and management in organizational contexts. This could include topics related to leveraging traditional to information systems approaches for capturing structured domain knowledge, such as through conceptual models. Knowledge representation is a fundamental and traditional part of research in information systems (Burton-Jones et al., 2017; Recker et al., 2021), making the IS community well-positioned to support RAG, ICL and other, similar initiatives.

Another research opportunity is supporting large language models with external systems that provide specific capabilities LLMs presently lack. Researchers and companies globally are rapidly responding to the opportunity to enhance the core of large language models with additional capabilities. For example, Lyu et al. (2023) propose using external solvers for math and reasoning tasks. Popular tools as well as development projects include systems that support query formulation, especially permitting drawing from personalized sources, such as corporate databases. Other tools offer reliable calculations (e.g., based on WolframAlpha API), and data visualizations (Zhuang et al., 2023). Tools emerge that integrate LLMs with external knowledge systems and platforms (e.g., travel websites permitting creation of complex itineraries based on highly personalized scenarios) (Zhang et al., 2024a, b)}. Promising research questions for information systems scholars include identification and exploration of opportunities for integrating LLMs with other, complementary IS systems. Information systems integration has been an important topic in IS, such as in the context of enterprise resource planning and social media. Another opportunity is the development of new tools that enhance LLMs, leveraging the vibrant design science IS tradition.

An emerging AI frontier is Large Action Models (LAMs), which seek to integrate the insights from LLMs with the action capabilities of autonomous agents, permitting intelligent action (Zhang et al., 2024a, b). This will allow, for example, one to directly book a vacation based on the complex itinerary created with the use of an LLM. In addition to the exploration of potential benefits of LAM, and development of LAM systems, giving direct autonomy to LLMs opens a host of questions of process, economics, trust and ethics of delegation of agency to AI.

Finally, generative AI advances the theory of information systems development by ushering in a new kind of information systems: *meta-information systems*. These are systems that could, in principle, take greater agency in IT development and monitoring, creation, and maintenance of other information systems. Indeed, GenAI already has capabilities that are important in information systems development; for example, the ability to analyze and structure textual documents. Hence, these tools can be used to augment and potentially automate requirements elicitation and analysis. The ability to produce database schemata and programming code can also be used to develop software components, such as database and application code and user interfaces. In this way, GenAI can become a type of meta-information technology; that is, a technology that permits the development and maintenance of other systems. Although we lack a theory of meta-information systems, GenAI could become an important use case.

### 4.5 Significant IS Research Topics

Technology has evolved significantly over time. In the history of technological transformation, agricultural mechanization liberated over 90% of farmers in developed countries. Similarly, robotic automation could release the majority of blue-collar workers from factory flowlines. Perhaps GenAI and associated business applications could release white-collar workers from office desks for many conventional processes and tasks, thereby changing the structure of their jobs and how they acquire knowledge (Alavi, 2024; Alavi & Westerman, 2023; Storey, 2025). If so, how can information systems researchers support this change? The research challenges and topics discussed in this paper are exemplary and the exploration of research frameworks, theories, and directions towards achieving this goal. Clearly, information systems researchers will strive to offer contributions to GenAI as it continues to evolve.

Researchers in information systems should help shape the impact of the progression of GenAI, focusing on its sociotechnical aspects and properties, as well as the manner in which it interacts with individuals, organizations, and society. This leads to various research challenges and opportunities.

- *Understand and improve business technology based on GenAI*. Design science researchers in information systems will have an opportunity to study technological issues in GenAI applications in business.
- *Understand the impacts of GenAI on individuals including workers and general users*. GenAI applications and plug-ins will be increasingly used by business work-ers and general users, which will require a renewal of many research topics in end user computing in the era of GenAI.

**Table 4** Research Opportunities on Generative AI for Information Systems

| Challenges | Research topics and examples |
| --- | --- |
| Understand and improve business technology based on GenAI | • Is generative AI transformative or disruptive to business? What are its application conditions during adoption?<br>• What representation(s) is needed (e.g., natural language, program code, algorithmic formulations) for the management of AI knowledge?<br>• What multimedia generation is appropriate? How should text, image, video, sound be synchronized based on tasks in business processes?<br>• Are there different types of GenAI applications for various users, e.g., natural language explanation (customers), algorithm specification (data scientists), and biases or unintended consequences (managers)? |
| Understand the impacts of GenAI on individuals including workers and general users | • How to evaluate and increase trust in GenAI and the systems with which it interacts?<br>• How can GenAI be used to augment existing knowledge bases?<br>• What are the potential negative impacts or the dark side of GenAI on human behavior, and what are the remedies?<br>• What are the implications for knowledge production and the amount of knowledge workers required to learn and retain?<br>• What is the importance of and emphasis needed for intelligent search?<br>• How can goodness and fairness of knowledge outcomes be assessed?<br>• Is it possible to identify and stop undesirable people from making connections via GenAI? If so, how? |
| Understand the impacts of GenAI on organizations in terms of processes and structures | • How is GenAI infused and adopted in organizations?<br>• What is the impact on business strategy?<br>• How do we manage and utilize the technology in organizations?<br>• What changes to managerial practices are needed?<br>• How do researchers model business tasks and processes that involve both human and machine agents?<br>• How can human and machine collaboration be optimized?<br>• What kind of human-in-the-loop or AI-human collaboration is required?<br>• What kind of interaction and supporting interfaces are needed? |
| Understand inter-organizational impacts of GenAI | • Can GenAI change the competitive situation and outcome of an industry; e.g., via more intelligent and dynamic pricing that might affect the market?<br>• How can GenAI improve productivity and efficiency; e.g., impact on organization's knowledge base, communications, R&D, product design?<br>• What are the potential adverse organizational impacts of GenAI, and what are the mitigation strategies? |
| Understand mission-critical business domains for GenAI adoption | • What is the impact of GenAI on mission-critical domains, such as medicine, military, transportation, food industry?<br>• What are implications for teaching information systems and other disciplines?<br>• How do we integrate GenAI into educational processes? |
| Understand legal and governance issues of GenAI | • What governance needs to be put into place?<br>• What if the collective influence of a company (companies) provides the majority of the input for connectionism features?<br>• How should internal and external risks be managed?<br>• How do we choose tasks while minimizing personal risk in exposure to penalty and lawsuits? |
| Understand broader societal issues of GenAI | • What are new ethical challenges of GenAI?<br>• What is the need for reference identification and verification of plagiarism?<br>• What is the extent of automation and job replacement? How do we measure it? Does it create "societal stress?"<br>• To what extent can creativity be achieved with GenAI, particularly for teams?<br>• How can we prevent personal information from being stolen in the era of GenAI? How can we protect individual's intellectual property while allowing data to be used in GenAI?<br>• Will GenAI change the labor markets and in what way? |

- *Understand the impacts of GenAI on organizations in terms of processes and structures*. GenAI will undoubtedly affect the capabilities of IT and individuals, leading to changes in business processes and structures.
- *Understand inter-organizational impacts*. When GenAI permeates organizations, the types of interactions between and among organizations will vary, leading to relationship and interactive changes.
- *Understand mission-critical business domains*. Research impacts can be more significant in mission-critical business domains such as healthcare and finance; therefore, special attention should be paid to these areas.
- *Understand legal and governance issues*. AI can lead to negative outcomes and leading to a dark-side of GenAI, requiring the studies of legal and governance issues.
- *Understand broader societal issues*. As GenAI is applied in various business sectors, additional societal issues, such as privacy and security, should be studied.

Table 4 provides examples of these research topics, which have a systems foundation. Additional possibilities will continue to emerge as this technology continues to be developed and applied in unique and interesting ways. The many research opportunities will offer a vast area of research for information systems scholars, whether they pursue technical, behavioral, managerial or economic research.

## 5 Conclusion

Recent advancement of GenAI is diffusive and penetrating because this technology is readily available and can be used with natural languages without a great deal of user training. As such, GenAI has the potential to affect more aspects of business operations than most previous technologies. In response to the explosion of applications, this paper has examined GenAI as the next generation of AI, which raises issues related to the role of technology (new, emerging, generative, or potentially transformative) in business and society. For information systems researchers, there are many important challenges that require careful consideration of both technical and societal aspects of GenAI. The paper has proposed an agenda for continued information systems research, identifying the potential contributions the field can make.

**Acknowledgements** J. Leon Zhao's work is partially supported by the National Natural Science Foundation of China grants No. 72031001 and 72231004. Roman Lukyanenko's work is supported by the University of Virginia. This research was also supported by the J. Mack Robinson College of Business, Georgia State University. The authors wish to thank the Editors-in-Chief, Dr. Ram Ramesh and Dr. H. Raghav Rao, as well as the anonymous reviewers for their careful consideration of this manuscript. Thanks also to Springer Production team for their efforts in producing the final version.

**Authors' Contributions** All authors contributed to the study conception and design. Material preparation, reference collection and manuscript writing were performed by all authors collaboratively. All authors read and approved the final manuscript.

**Data Availability** There is no data repository beyond the content of this manuscript.

## Declarations

**Veda C. Storey** is a Distinguished University Professor and the Tull Professor of Computer Information Systems and professor of computer science at the J. Mack Robinson College of Business, Georgia State University. Her research interests are in data management, conceptual modeling, and design science research. She is particularly interested in the assessment of the impact of new technologies on business and society from a data management perspective. Dr. Storey is a member of the AIS College of Senior Scholars and the steering committees of the International Conference of Conceptual Modeling and the Workshop on Information Technologies and Systems. She is a recipient of the Peter P. Chen Award, an ER Fellow, an AIS Fellow, and an INFORMS Fellow.

**Wei Thoo Yue** is a Professor of Management Information Systems in the Department of Information Systems at City University of Hong Kong. He received his Ph.D. in Management Information Systems from Purdue University. Prior to joining City University of Hong Kong, he was a faculty member at the University of Texas, Dallas. His research interests focus on the economics of information systems. His work has appeared in Management Science, Information Systems Research, MIS Quarterly, Journal of Management Information Systems, Decision Support Systems, and other journals. He currently serves as Senior Editor for Production and Operations Management.

**J. Leon Zhao** is a Presidential Chair Professor, Director of Center on Blockchain and Intelligent Technology, Co-head of Information Systems and Operations Management, School of Management and Economics, Chinese University of Hong Kong, Shenzhen. He was a chair professor at City University of Hong Kong and Eller Professor at University of Arizona, respectively. He has edited over 20 special issues for academic journals including MIS Quarterly, Information Systems Research, and Journal of Operations Management. He received an IBM Faculty Award in 2005, and National Chang Jiang Scholar Chair Professorship first at Tsinghua University in 2009 and again at the Chinese University of Hong Kong, Shenzhen in 2022.

**Roman Lukyanenko** is an associate professor at the McIntire School of Commerce, University of Virginia. His research interests include data management and research methods (validity and artificial intelligence in literature reviews). Roman actively develops ideas, tools, and methods to improve data management and research practices. These solutions received major awards, including INFORMS Design Science Award, Governor General of Canada Gold Medal, Hebert A. Simon Design Science Award. Roman's research appeared in Nature, MIS Quarterly, Information Systems Research, ACM Computing Surveys. His 2019 paper on quality of crowdsourced data received the Best Paper Award at MIS Quarterly.